\documentclass[10pt,journal,compsoc]{IEEEtran}

\usepackage{atbegshi}
\AtBeginDocument{\AtBeginShipoutNext{\AtBeginShipoutDiscard}}

\ifCLASSINFOpdf
  \usepackage[pdftex]{graphicx}
\else
   \usepackage[dvips]{graphicx}
\fi
\usepackage{breqn}

\usepackage{listings}

\usepackage{array}
\usepackage{url}

\usepackage[font=small,skip=0pt]{caption}

\usepackage{booktabs} 
\usepackage[table,xcdraw]{xcolor}

\usepackage[export]{adjustbox}

\usepackage{xcolor}
\usepackage{soul}
\usepackage[colorinlistoftodos]{todonotes}

\usepackage{hyperref}

\begin{document}
%
\title{Dynamically Reconfigurable Variable-precision Sparse-Dense Matrix Acceleration in Tensorflow Lite}
%
%
%
%

\author{Jose Nunez-Yanez, Andres Otero, Eduardo de la Torre}
\IEEEcompsocitemizethanks{\IEEEcompsocthanksitem J. Nunez-Yanez 
of Electrical Engineering, University of Linköping, Linköping, Sweden. 
E-mail: jose.nunez-yanez@liu.se\IEEEcompsocthanksitem Andres Otero, Eduardo de la Torre, Centro de Electronica Industrial, UPM, Madrid, Spain. 
E-mail: {joseandres.otero,eduardo.delatorre}@upm.es}

\thanks{Manuscript received April 19, 2005; revised August 26, 2015.}

\IEEEtitleabstractindextext{%
\begin{abstract}

In this paper, we present a dynamically reconfigurable hardware accelerator called FADES (Fused Architecture for DEnse and Sparse matrices). The FADES design offers multiple configuration options that trade off parallelism and complexity using a dataflow model to create four stages that read, compute, scale and write results. FADES is mapped to the programmable logic (PL) and integrated with the TensorFlow Lite inference engine running on the processing system (PS) of a heterogeneous SoC device. The accelerator is used to compute the tensor operations, while the dynamically reconfigurable approach can be used to switch precision between \textit{int8} and \textit{float} modes. This dynamic reconfiguration enables better performance by allowing more cores to be mapped to the resource-constrained device and lower power consumption compared with supporting both arithmetic precisions simultaneously. We compare the proposed hardware with a high-performance systolic architecture for dense matrices obtaining 25\% better performance in dense mode with half the DSP blocks in the same technology. In sparse mode, we show that the core can outperform dense mode even at low sparsity levels, and a single-core achieves up to 20x acceleration over the software-optimized NEON RUY library.   

\end{abstract}

}

\maketitle

\IEEEdisplaynontitleabstractindextext

%
\IEEEpeerreviewmaketitle

\IEEEraisesectionheading{\section{Introduction}\label{sec:introduction}}

In this research, we present FADES (Fused Architecture for DEnse and Sparse tensor processing) dataflow engine and its extension with dynamically reconfigurable (i.e. Xilinx DFX) capabilities to support floating-point and 8-bit precision arithmetic. DFX enables the modification of blocks of logic by downloading partial bit files that change the functionality on-the-fly without interrupting the operation of the rest of the system. DFX makes more efficient use of the silicon, allowing designers to move to smaller devices and reduce power. FADES is integrated as an accelerator for TensorFlow Lite (TFLite) that is TensorFlow's lightweight solution for mobile and embedded devices suitable for edge deployment promoted by Google. This integration means that FADES benefits from extensive prior development research done by the TensorFlow Lite community on sophisticated quantization-aware training and pruning ~\cite{tflite} and directly replaces the high-performance matrix multiplication library RUY \footnote{\url{https://github.com/google/ruy}}. A hardware accelerator that can perform both floating-point and 8-bit arithmetic is useful because TFLite natively supports both precision data types. Deep learning models are continuously evolving, so there is a need for flexible hardware solutions that can scale and adapt to increasing network complexities and diversity. Motivated by these observations, the contributions of this paper are as follows:

\begin{itemize}

\item We present the FADES variable-precision sparse-dense dataflow accelerator designed using a high-level synthesis approach and its integration into the Tensorflow Lite inference engine.
\item We develop 8-bit integer and floating-point configurations in sparse and dense matrix processing modes and compare the performance with a systolic array hardware accelerator and optimised software alternatives.
\item We propose a design methodology that combines high-level synthesis and dynamic function exchange to generate and deploy variable-precision neural accelerators.
\item We release the designs open-source to promote further work in this field at: \footnote{\\url{https://github.com/eejlny/gemm_spmm}}
\end{itemize}

This paper is organized as follows: section 2 reviews state-of-the-art solutions for edge deployment, showcasing that current neural network hardware focuses on 8-bit precision, application specific architectures and dense matrix operations. Section 3 motivates the significance of this work highlighting which technical attributes are different from current available hardware. Section 4 describes the proposed hardware architecture for high-performance dense and sparse tensor operations with variable-precision float/int support. Section 5 evaluates the performance mapped to a target edge device - the Zynq UltraScale+ MPSoC comparing it with a systolic design, the optimized ARM library for low-precision tensor arithmetic RUY, and a reference C++ implementation. Section 6 presents the dynamically reconfigurable methodology and its integration with software-defined design frameworks and the TensorFlow Lite inference engine. Section 7 validates performance, power and energy consumption using different accelerator configurations integrated in TFLite and other state-of-the-art hardware targeting comparable edge devices. Finally, section 8 concludes this paper and proposes future work. 

\section{Related work and Motivation}\label{sec:background}

\subsection{Flexible-model Neural accelerators}

In this section we discus neural accelerators that can support multiple models without hardware changes. Efforts in accelerating the execution of TensorFlow Lite models using flexible hardware have focused on deploying systolic architectures for tensor operations to obtain very high throughput. For example, Google offers a low-cost and low-power version of the Google TPU called EdgeTPU~\cite{edgetpu}, which can run dedicated neural networks with 8-bit precision. The systolic array size in the EdgeTPU has 64 x 64 multiply-add cells obtaining 4 TOPS at 480MHz, and it is much smaller than the TPU cloud configurations. Layers with floating-point precision will run on a external CPU that would act as the host for the EdgeTPU device. Xilinx has also focused on inference including support for TensorFlow, with the Xilinx DPU~\cite{dpu} unit. It is composed of a register configuration unit, the data controller and convolution computing modules optimized for the FPGA hardware resources. The original hardware is specialized for convolutional neural networks, although alternative architectures and hardware configurations are made available for other model types. The DPU hardware supports integer precisions of 4-bit, 8-bit and 16-bit. Xilinx DPU architectures are generally specialized for particular model types, and a vendor-specific compiler, optimizer and quantizer are needed in order to deploy TensorFlow models in the supported devices. The approach in this work is closer to the EdgeTPU than the Xilinx DPU since it can be applied to any TFLite model without any additional compilation or optimizations. While the EdgeTPU supports 8-bit precision exclusively, FADES supports integer precisions of 1, 2, 4, 8-bit and single and half floating-point. In this paper, we focus on 8-bit and single floating-point precisions, which are compatible with current versions of the TensorFlow Lite inference engine that runs on the host processor.

\subsection{Model-specific neural architectures}

In this section we review hardware architectures that requires changes if a new model is to be deployed. There is a significant body of research that has focused on creating tools and architectures targeting particular network models, such as simple FCNNs (fully connected neural networks)~\cite{fcnn}, CNNs (convolutional neural networks)\cite{cnn} or LSTMs (long short-term memory networks)~\cite{lstm}. For example, in ~\cite{brds}, a specialized architecture for LSTM-type networks on FPGAs with sparsified weights is presented. It works by determining at design-time the number of PEs (processing elements) that are required for each row of the sparse-matrix vector operation. A dual-pruning strategy is proposed for the different weight matrices involved in LSTM-type networks. The LSTM-type network is also addressed in \cite{stocastic} integrating stochastic computing principles in the LSTM model. The hardware that uses the stochastic number representation with multiplication arithmetic done with XNOR gates and additional with mux or parallel counters. The FPGA demonstratoe shows significant lower power usage than reference hardware with conventional binary arithmetic  without affecting overall accuracy. Examples of architecture specialization in FPGAs also include FINN~\cite{finn} and hls4ml~\cite{hls4m} - two well-developed frameworks supporting arbitrary precision hardware. hls4lm has been shown to achieve very high performance on very low latency problems such as particle colliders. It uses quantization-aware training and pruning based on the QKeras library and exports the resulting configuration as a C++ Vivado HLS description ready to be implemented in an FPGA. The proposed example neural network in ~\cite{hls4m} consists of 3 dense layers and a softmax layer with a precision of 14 bits with 6 integer bits. Pruning forces many weights set to zero and removes the hardware resources associated with these weights. Pruning does not affect latency because the non-zero weights define the longest path. Therefore the depth of the network remains unchanged. Other task-specific neural network frameworks optimized for FPGA mapping include LUTnet~\cite{lutnet} and LogicNets~\cite{logicnet}. These frameworks advance the concept of using the FPGA LUTs to implement 2-input XNORs between weights and activations, as used in binary networks, in order to exploit the capabilities of the multi-input LUTs. The weights are baked into the logic function implemented in the LUT, and a variable number of LUT inputs are used for the binary activations. The number of inputs needed in the LUT is reduced with pruning techniques, resulting in better performance and lower logic complexity. In any case, specialization means that the architecture and final implementation are task-specific. In ~\cite{survey} different hardware architectures to accelerate CNNs in FPGAs are discussed, indicating how CPU+FPGA approaches can benefit from mapping to the FPGA only the parts where the FPGA is very efficient while more irregular computation is done on the CPU. This is the heterogeneous approach selected in this paper that can also support large models without requiring large FPGA devices. Heterogeneous devices combine different processing resources such  as CPUs, FPGAs and GPUs enabling low-power edge-intelligence applications with improved integration such as the low-latency hybrid data initialization proposed in \cite{integration} that improves how input and parameter network data are loaded by the different devices. Accelerators focusing on sparse computation include ~\cite{efficient} that shows similarities with our work  with the sparse weights stored using CSR format, shape-wise pruning and an equivalent ultrascale device although the development is done in Verilog and not HLS. The accelerator achieves 990 GOps/second on VGG16 with a DSP efficiency of  0.367 GOPS/DSP and uses different architectures depending on the CNN model. Heterogeneous devices have also been used for dense matrix processing and in ~\cite{deeper} a tool flow is proposed for dynamic precision quantization with 8 and 4-bit values for the weight matrix with a variable number of PEs in each layer depending on the layer precision. The design uses a classical line buffer optimized for the VGG network with a fix size window with 3x3 values. Our approach is not specialized for particular filter sizes and it is design to process any matrix shape used by the TFlite model format. 

\section{Research Motivation}

In contrast to the research presented in section 2.2, the work presented in this paper avoids creating a specific architecture for a particular neural network model. Instead, it aims to be a direct hardware-based drop-in replacement of a software matrix multiplication library such as Google RUY and it is conceptually similar to the Google EdgeTPU. The benefit is that while the EdgeTPU uses a systolic architecture and supports only dense matrix multiplication with 8-bit precision, FADES can work with multiple precisions and dense/sparse models efficiently, minimizing data stalls by streaming data into internal FIFOs and BRAMs and performing irregular data accesses only to the BRAM internal memory. FADES consists of an ensemble of dynamically reconfigurable accelerator overlays that work with models optimized using the pruning and quantization techniques part of the TensorFlow Model Optimization extensions Toolkit ~\cite{mot}. In this way, we benefit from the extensive amount of prior research done by the TensorFlow community on sophisticated quantization-aware training and pruning. Therefore, the objective of the paper is not to propose novel training techniques to compensate for the possible degradation of accuracy due to sparsity. In our previous work~\cite{array}, we investigated the accuracy effects of arbitrary quantization with LSTM and CNN layers, and we presented hardware for independent sparse and dense computations using small Xilinx Zynq SoC devices. The sparse architecture was an extension of matrix-vector hardware, which limited parallelism, and there was no specific support for TensorFlow Lite. 
In this paper, we target the larger Zynq Ultrascale devices with a novel fused architecture that reuses hardware resources and is integrated into the inference engine of TFLite. We deploy dynamic reconfiguration with high-level synthesis to support floating-point and integer precisions. The accelerator performance and power consumption are compared with virtual reconfiguration alternatives and state-of-the-art software and hardware acceleration libraries. Table \ref{tab:compare} compares selected technical attributes with the FADES proposal to highlight differences.

\begin{table}[!htb]
\resizebox{\columnwidth}{!}{%
\large
\begin{tabular}{ccccccc}
      Accelerator & \shortstack {Model set \\ at run-time}  & \shortstack {Target \\ technology}   & \shortstack {Dynamic \\ float-int \\ precision} & \shortstack {Sparse \\ support} & Flexibility & \shortstack {Tensorflow \\ integration}\\
\hline
EdgeTPU \cite{edgetpu} & Yes & ASIC & No & No & Low & Yes \\
\hline
DPU \cite{dpu} & Yes & FPGA & No & Yes & Low & Yes  \\
\hline
FINN  \cite{finn} & No & FPGA & No & No & High & No \\
\hline
HLS4ML \cite{hls4m} & No  & FPGA & No & No & High & No \\
\hline
LUTNET \cite{lutnet} & No & FPGA & No & No & High & No \\
\hline
\cite{deeper} & No & FPGA  & No & No & High & No \\
\hline
\cite{efficient} & No & FPGA  & No & Yes & High & No \\
\hline
FADES & Yes & FPGA & Yes & Yes & High & Yes
\end{tabular}%
}
\caption{Hardware comparison}
\label{tab:compare}
\end{table}

\section{Fused architecture}\label{sec:fused}

\subsection{System architecture}
\begin{figure}[htb]
  \includegraphics[width=0.8\columnwidth,center]{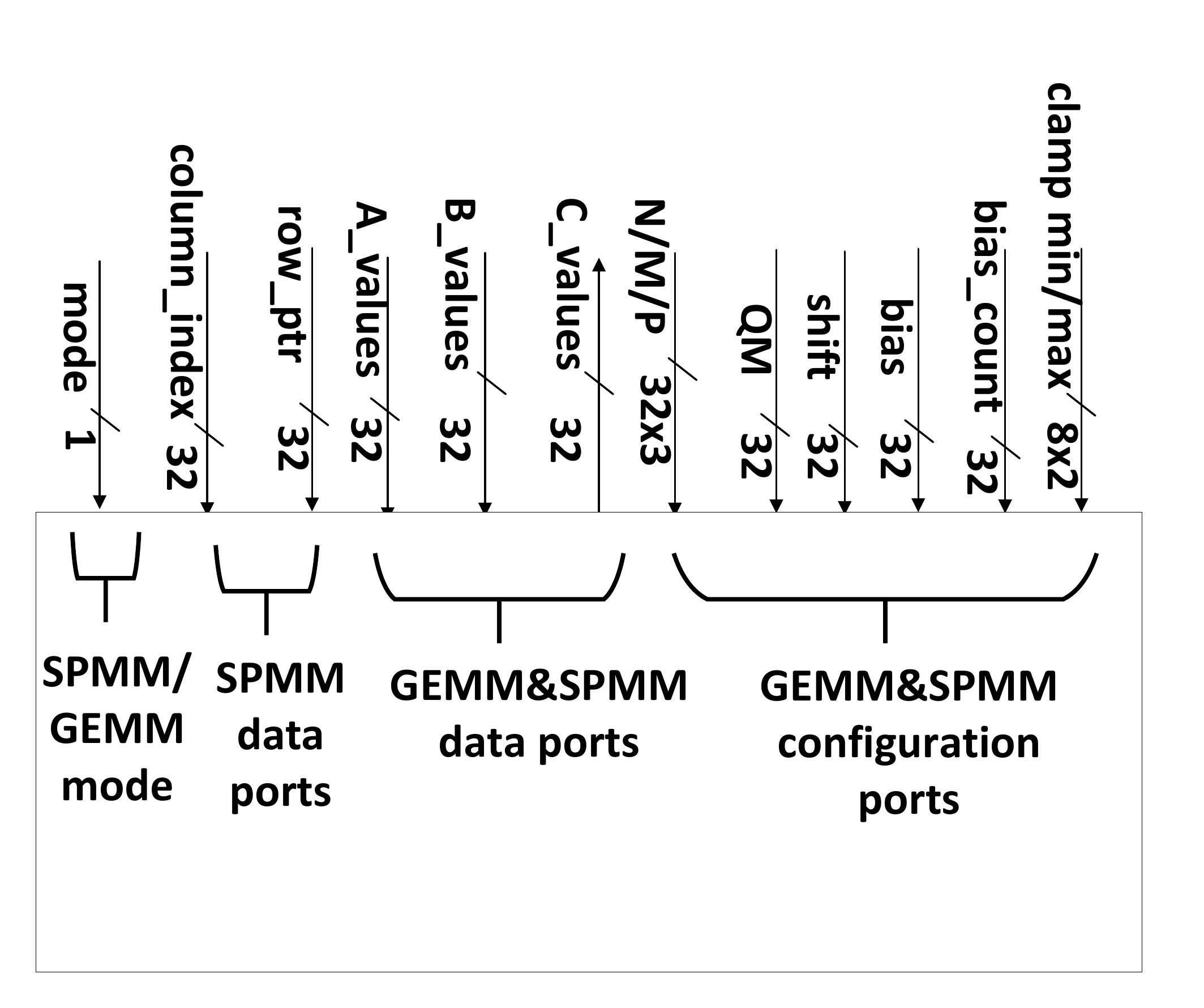}
  \caption{Top-level block diagram}
  \label{fig:block_diagram}
\end{figure}

The use of low precision data types and sparse neural representations are two research trends that favour the adoption of FPGAs in deep learning. Motivated by these trends, we design the FADES dataflow architecture to accelerate TFLite matrix operations with support for asymmetric quantized activations, column-major matrix write, per-filter\slash per-axis bias values and the current scaling TensorFlow Lite specifications. FADES computes the tensor operations of the DNN models, representing up to 95\% of the execution time for the tested neural networks. In FADES, we design the SPMM (SParse Matrix Matrix) and GEMM (GEneral Matrix Matrix) accelerators using a tile approach to deal with a $B$ matrix composed of a large number of columns. We use a high-level synthesis (HLS) description to define the architecture. The challenge of supporting both GEMM and SPMM simultaneously  is how to avoid unnecessary hardware replication and performance degradation in the HLS description. A top-level block diagram showing the architecture ports is shown in  ~\ref{fig:block_diagram}. The mode input sets the hardware in sparse or dense modes, the data ports move the data values for for the input A,B and output C matrices and the configuration ports control Tensorflow Lite int8 scaling and clamping parameters, optional bias and matrix dimensions. The matrix shape being processed is defined at run-time with parameters \textit{$N=A$ rows}, \textit{$M=A$ columns} and \textit{$P=B$ columns}.

\begin{figure*} [!htbp]
  \includegraphics[width=18cm,height=13cm]{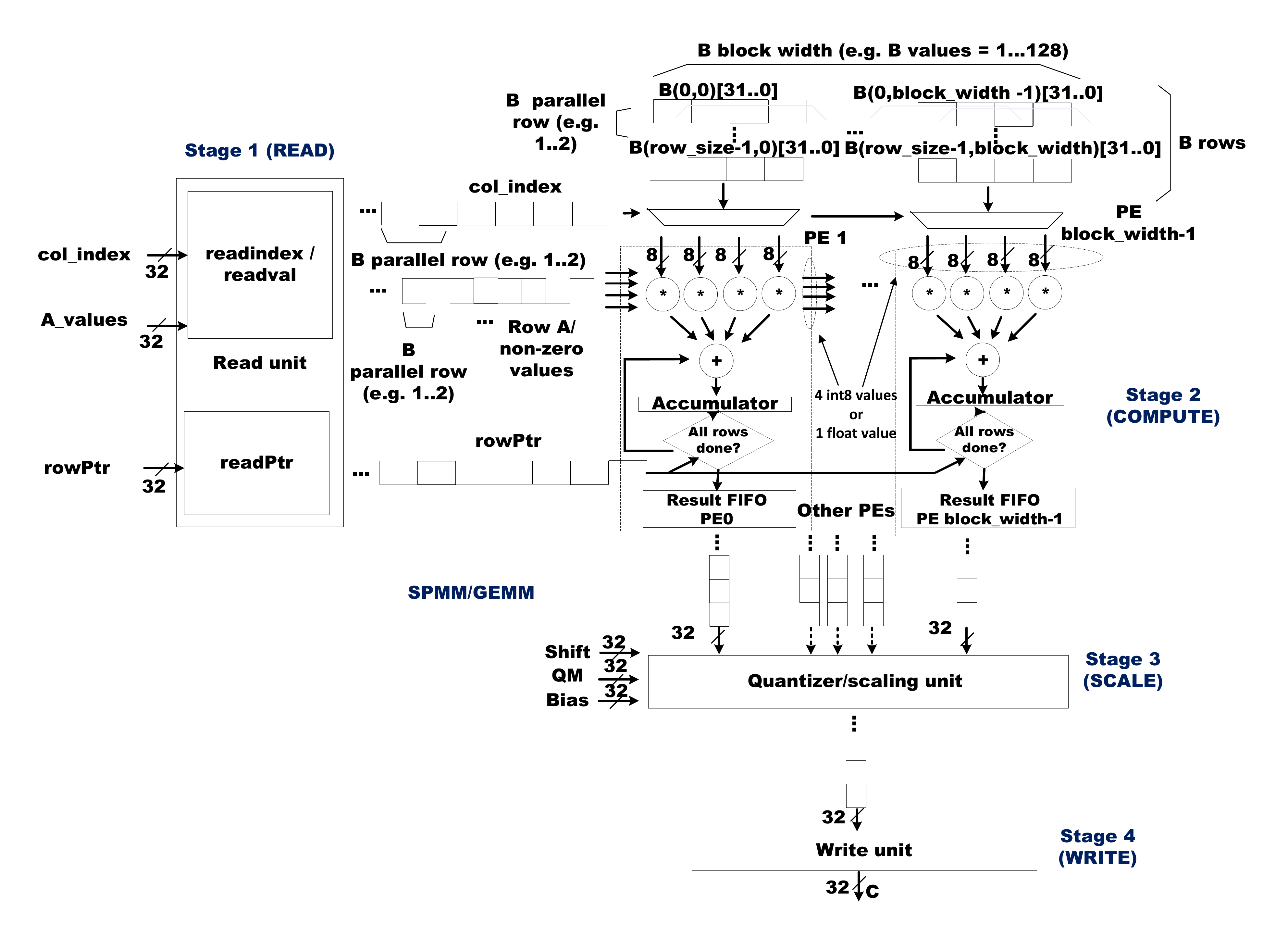}
  \caption{Fused GEMM/SPMM architectural components}
  \label{fig:kernel}
\end{figure*}

Figure  ~\ref{fig:kernel} shows the internal organization of the dataflow consisting of 4 stages interconnected with streaming FIFOs for both GEMM and SPMM. All the dataflow stages are coded in HLS to obtain an efficient initialization interval (e.g. II) of 1. An initialization interval of 1 in high-level-synthesis means that a new iteration of the loops can be started in each clock cycle which is a critical requirement for high performance. In FADES, the streaming architecture means that a tile of the dense matrix is initially buffered inside the FPGA device, then the sparse matrix in CSR format is streamed from memory into the accelerator. As the sparse values are streamed, they are computed with multiple values of the dense matrix depending on the tile size. After the tile is processed, a new tile is loaded, and the streaming of the sparse data is performed again. This approach means that all irregular accesses are done to local BRAM memory and do not result in stalls or additional latency.

Stage 1 is responsible for loading a $B$ matrix block with a column count that equals $PEs*4$ bytes (each PE processes four 8-bit elements or one single-precision floating-point element) and row count that equals the number of rows in the $B$ matrix. It then starts streaming elements of the $A$ matrix in raw format for GEMM, and in CSR (Compressed Sparse Row) format with \textit{column\_index} and \textit{non-zero} values for SPMM. This means that in SPMM mode, the accelerator performs a variable number of reads of weight values and column indices per matrix. In GEMM mode, the number of reads is constant and defined by the row size multiplied by the number of rows in the weight matrix, while in SPMM is defined by the number of non-zero elements present in the weight matrix. This does not represent a performance limiting factor in the dataflow architecture because the READ stage is independent of the other stages, and it can run at full speed over all the data elements present in the  $A$ matrix for both GEMM and SPMM as long as there no buffer underflows or overflows. 

Stage 2 is the main computing loop that instantiates the bulk of DSP blocks. It aims at activating all PEs in parallel in each clock cycle. This is the case for all $B$ tiles with a number of columns equal to or larger than the number of PEs specified by $B\_WIDTH\_BLOCK$. $B\_BLOCK\_WIDTH$ is the main parameter related to parallelism and it defines the number of PEs that work in columns of the activation matrix $B$ in parallel. Typically, the last tile contains a number of columns lower than $B\_WIDTH\_BLOCK$, and in this scenario, some of the PEs do not write their output FIFOs. This enables arbitrary matrix shape support that is not a multiple of the tile size. 

Stage 3 uses the values of $QM$ (quantization multiplier) and $shift$ to implement the scaling strategy as per Tensorflow Lite specifications for int8 values. It reads the raw values produced by stage 2 and writes the corresponding scaled values to stage 4. Stage 3 forwards directly raw values to stage 4 in float mode. Finally, stage 4 reads these values and writes them to the correct memory addresses to properly construct the output matrix C assembling the multiple tile outputs. 

The accelerator offers multiple compile-time configuration options that are summarized in table~\ref{tab:options}. The columns in the table represent the number of cores (CCs), the number of processing elements per core (PEs), the number of parallel rows per core (PRs) that sets the number of rows of A that are processed in parallel, the supported modes (SPMM only, GEMM only or FUSED that enables both SPMM and GEMM), transpose output matrix enabled/disabled (TRANS) and if scaling is enabled/disabled (SCALE). SCALE ensures that the results follow the TFLite specification using per-filter quantized multiplier parameters. Per-filter quantized multiplier parameters are part of the recent TFLite versions, and they enable different scaling parameters for each filter to improve accuracy. TRANS ensures that the result matrix is written in a column-major format, as it is currently required by the TensorFlow Lite inference engine.  The CC option instantiates multiple independent cores in the device and splits the computation evenly among them. The splitting can be done either with multiple blocks of the $A$ matrix and a single $B$ matrix, or a single $A$ matrix and multiple blocks of the $B$ matrix. The first option divides the sparse matrix $A$ into multiple blocks, and it is the option used in this paper to optimize the read of $A$ values and column indexes over multiple ports.

\begin{table}[!htb]
\resizebox{\columnwidth}{!}{%
\begin{tabular}{llllllll}
     CCs & PEs  & PRs   & SPMM & GEMM & TRANS & SCALE \\
\hline
    1 to 4 & 2 to 256 & 1 to 2 & 0 to 1 & 0 to 1 & 0 to 1 & 0 to 1
\end{tabular}%
}
\caption{Accelerator  configuration options}
\label{tab:options}
\end{table}

\subsection{Float-int8 hardware support}

In the FADES design each data word is 32-bit wide and it is formed with weight/activation values whose width depends on the required precision. In our previous work, we show how the 32-bit value can be interpreted as multiple 2, 4 or 8-bit int values~\cite{array}. In this paper, we focus on analyzing the \textit{int8} and float modes that are directly compatible with the TFLite inference engine running on the PS. Although it is generally accepted that 8-bit precision is enough for deep-learning inference, floating-point precision may be required by some layers within the network or in certain types of networks such as recurrent neural networks ~\cite{pooja}. Floating-point precision is also useful when network re-training and not just inference is required in the application to follow changes in the input data statistical properties. Code snippets for the compute function are shown in listing 1 and 2 for the \textit{float} and \textit{int8} precisions, respectively. Note that the $int8$ listing shows a partial accumulation of the four results obtained in one PE, while the float listing does not do this accumulation since there is only one result per PE. In both cases, a not-shown outer loop performs the accumulation of results over all the rows of the activation matrix $B$.

\begin{lstlisting}[xleftmargin=0.5cm,numbers=left,language=Python,breaklines=true,postbreak=\mbox{\textcolor{black}{$\hookrightarrow$}\space},caption=Float compute kernel example]

for (int j = 0; j < B_WIDTH_BLOCK; j++) 
{
    #pragma HLS UNROLL
    #pragma HLS PIPELINE
    ap_int<32> b_block_int = b_block[b_row][j];
    ap_uint<32> A_val = a_value;
    FTYPE a_val_float=*(FTYPE*)&A_val;
    FTYPE b_val_float=*(FTYPE*)&b_block_int;
    float rhs_float = float(zero_point_rhs);
    acc[j] = (a_val_float)*(b_val_float-rhs_float);
} // j loop


\end{lstlisting}

\begin{lstlisting}[xleftmargin=0.5cm,numbers=left,language=Python, ,breaklines=true,postbreak=\mbox{\textcolor{black}{$\hookrightarrow$}\space},caption=Int8 main compute kernel]

for (int j = 0; j < B_WIDTH_BLOCK; j++) {
    #pragma HLS UNROLL
    #pragma HLS PIPELINE
    for(int z = 0; z < DTYPE_LENGTH; z+=8) {
        ap_int<8> A_val = a_value.range(z+7,z);
        ap_int<8> B_val = b_block[b_row][j].range(z+7,z);
        acc[j] += A_val*(B_val-zero_point_rhs);
    }
} // j loop



\end{lstlisting}

To maximize hardware reuse, the code is created to minimize differences that depend on the data precision. Data types in the interface and buffer memory are mainly defined as \textit{uint32} and only in the compute kernel pointer reinterpretation is used to interpret the \textit{uint32} value as four \textit{int8} components or as a single floating point library as conceptually shown in lines 7 and 8 in Listing 1 where \textit{A\_val} is a \textit{uint32} value and \textit{a\_val\_float} a float value. The operations done on \textit{a\_val\_float} will be standard floating point operations.

To obtain high performance in float mode it is not enough to just interpret the 32-bit word as a floating-point value. The problem is that the latency of a floating-point ADD (FADD) used for accumulation in the Zynq Ultrascale is higher than one. Consequently, to achieve one addition per clock cycle, the pipeline inside FADD must be filled with interleaved operations. On the other hand, the latency for the integer addition is one clock cycle, so the C synthesis tool can achieve the initialization interval of one without interleaved operations. In order to achieve an equivalent initiation interval of one in the float kernel, we interleave FADD\_LATENCY partial accumulations onto the same core, each completing every FADD\_LATENCY cycles. The C synthesis tool recognizes that it can schedule the partial accumulations onto a single adder core on alternating cycles. The optimal value of FADD\_LATENCY increases with the target frequency and has a value of 6 at 200 MHz which is the frequency selected for all the implementations running in the Zynq UltraScale+ device. The resulting architecture uses 4 DSP48 blocks per 32-bit float operator and 1 DSP48 block per \textit{int8} operator resulting in a similar number of DSP48 blocks for the 32-bit float and 32-bit int8 configuration. Table~\ref{tab:complex} shows the complexity of different configurations as follows. The first number in the pair represents the number of cores, being 1 always in this case. The second one, is the number of processing elements, which can be 32 or 128. It must be noticed that the number of the DSP blocks in the float and \textit{int8} versions have some variation due to to the presence of the scaling module in the \textit{int8} configuration and the addition of interleaved partial results in the float configuration. The third row in this table shows the design complexity supporting int8 and float configurations simultaneously using multiplexers to select if the float or \textit{int8} datapaths are active. The pointer reinterpretation in the source code maintains the code for both configurations similar and maximizes the amount of logic reuse. In any case, the internal organization of the DSP48 blocks for 1 float and 4 int8 operations is different, so when both configurations are enabled, the compiler needs to duplicate the DSP48 usage. This additional resource utilization means that, for example, the 128 float-int8 configuration fails to meet timing at 200 MHz, although the individual configurations meet timing without issues. The main advantage of dynamic function exchange is that both configurations do not need to be deployed simultaneously. In section 6 we will explore how dynamic function exchange reduces resource usage and improves power consumption (due to that only the logic elements that participate in the computation need to be configured) and performance (since a core with more processing elements can be created without having to lower the clock frequency). 

\begin{table}[!htb]
\resizebox{\columnwidth}{!}{%
\begin{tabular}{lllllll}
      Configuration & LUTs(K)  & FFs(K)   & BRAM\_18Ks & DSP48Es\\
\hline
(1,32) float  &  90.9 &  120.5 &  186 &  181 \\
\hline
(1,32) int8  & 50.8 & 56.9 & 189 & 161 \\
\hline
(1,32) float-int8 & 96.8 & 121.6 & 187.5 & 325 \\
\hline
(1,128) float  & 198.2 & 245.9 & 552 & 661 \\
\hline
(1,128) int8  & 67.2 & 78.7 & 571 & 545 \\
\hline
(1,128) float-int8 & 221.8 & 260.8 & 564 & 1189 \\
\hline
(32x32) systolic int8 &  68.2K & 27K & 1041 & 1031 \\
\end{tabular}%
}
\caption{configuration complexity comparison}
\label{tab:complex}
\end{table}


In summary, the whole architecture is based on a DATAFLOW of the different functions or stages. Then, each stage uses pipelining that can be considered a fine-grained dataflow (or equivalently DATAFLOW can be considered a coarse-grained PIPELINE). The pipelining of the FADD accumulation is problematic when the operators are floating point values due to the latency of floating point add that is longer than one so the interleaving technique described in this section is used to achieved an overall latency of one for FADDs. The side effect is  that the floating point and integer logic are significantly different and dynamic reconfiguration will be used to switch between modes. 

\section{Initial Performance and functional validation} \label{initial}

\subsection{Experimental setup}

In this section, we validate the performance characteristics of the FADES core compared with high-performance systolic hardware ~\cite{systolic}. This initial performance evaluation considers a stand-alone implementation before its integration as part of the TFLite framework so a fair comparison with ~\cite{systolic} can be done while section \ref{full_perf} will analyze the performance obtained after TFlite integration. Xilinx SDSoC 2018.3 is used for both ~\cite{systolic} and the proposed design with a clock frequency of 200 MHz for the programmable logic while the processing system runs at the standard frequency of 600 MHz. To validate the correct functionality of the designs we have created a C test bench compiled with the RUY multiplication library. The matrices are initialized with random data and output generated by the hardware are compared with the software results produced by the RUY library to ensure correctness. The systolic architecture has been implemented on the same ZCU102 board equipped with the Zynq UltraScale device running Ubuntu 16.04. The systolic core has 32x32 = 1024 PEs, and each PE uses one DSP block as per the original design. We also consider as a comparison point the high-performance multiplication library RUY \footnote{\url{https://github.com/google/ruy}}, developed by Google engineers, that focuses on covering the matrix multiplication needs of neural network inference engines. RUY is used in TFLite, as an option, on the ARM CPU architecture. It is designed to achieve high performance not just on very large matrix sizes but on the sizes and shapes of matrices relevant for TensorFlow Lite applications. RUY fully uses assembly code to make use of ARM NEON and SIMD optimizations, and it supports both floating-point and 8-bit integer quantized matrices. It replaces EIGEN and GEMMLOWP, achieving better performance. An alternative to RUY is XNNPACK, but XNNPACK focuses exclusively on float operations, so it is not applicable in this work. XNNPACK also relies on the armv8.2 instruction set that is not available in armv8 processors such as the Cortex-A53, available in the Zynq UltraScale+ MPSoC used in this work. The systolic hardware only supports int8, while RUY can be configured to work with floats and int8 precision. We initially consider the dense mode with square matrices since neither systolic nor RUY can run in sparse mode. 

\subsection{Initial performance results}

Table~\ref{tab:square_performance} shows these results including a reference C++ implementation after validating the correct output of each test point. The table shows that FADES with a (1,128) configuration in int8 mode clearly outperforms the systolic hardware and RUY, especially  as the size of the matrix being processed increases. This is significant since the systolic configuration contains 32x32 cells with twice the amount of DSP blocks. It is also clear that the C++ implementation is much slower than the software optimized RUY. Overall, with a matrix size of 1024x1024 FADES (128,1)/int8 is ~18x faster than RUY in int8 mode and FADES (128,1)/float  is ~9x faster in float mode.  Notice that the FADES accelerator includes the per-filter scaling hardware needed in TFLite, so direct comparisons with other GEMM hardware that simply performs matrix multiplication must take this into account.  

\begin{table}[htb]
\resizebox{\columnwidth}{!}{%
\begin{tabular}{lllll}
    & 128 & 256 & 512 & 1024   \\
\hline
Systolic int8 & 0.29 & 1.27 & 5.34  & 21.92 \\
\hline
RUY int8 & 1.53 & 6.14 & 41.03 &  302.31 \\
\hline
C++ int8 & 547.4 & 4k & 35k  & 314k  \\
\hline
1,128 int8 & 0.307 & 1.001 & 3.618 & 16.58   \\
\hline
1,32 int8 & 0.365 & 1.575 & 8.65  & 54.78   \\
\hline
RUY float  & 1.94 & 14.73 & 64.63 & 500.02  \\
\hline
C++ float  & 583.48 & 4651.59 & 38923.8 & 423516  \\
\hline
1,128 float  & 0.089 & 3.28 & 12.88 & 56.23 \\
\hline
1,32 float  & 0.921  & 4.761 & 29.33 & 200.5 \\
\end{tabular}%
}
\caption{float/int8 performance with square matrices (ms)}
\label{tab:square_performance}
\end{table}

The (1,32) configuration uses 4x fewer DSP blocks and it is approximately 4x slower than the (1,128) configuration for the larger matrices, although this performance difference is diluted for the smaller matrices as expected. The lower resource usage for the (1,32) configuration enables more cores to be mapped to the Zynq Ultrascale device with the same DSP usage. Table~\ref{tab:multi_core} shows a performance comparison for a large matrix comparing the (1,128) solution (one core with 128 PEs) and the (4,32) solution (4 cores with 32 PEs each), and using the sparse mode with 3 levels of sparsity: low (~50\% sparse), medium (~70\% sparse) and high (~90\% sparse). Note that configurations (1,128) and (4,32) contain the same number of processing elements (i.e. 1x128 = 4x32 = 128 PEs). It is clear that for dense and low levels of sparsity, the multi-core configuration does not offer significantly better performance. The reason is that for dense matrices, the compute intensity is high, and the bottleneck is the compute power defined by the number of PEs that is equivalent in both configurations. On the other hand, as the sparsity increases, the performance bottleneck moves to the data movement, and the additional ports available in the multi-core configuration deliver noticeably better performance. The smaller cores used in the (1,32) configuration also favour introducing a DFX strategy compared with reconfiguring the whole device. The performance and power advantage of the DFX methodology using (1,32) cores is explored in the following sections.   

\begin{table}[htb]
\resizebox{\columnwidth}{!}{%
\begin{tabular}{lllll}
    & dense(ms) & sparse low & sparse medium & sparse high  \\
\hline
1,128,1 int8  & 32.776 & 19.75 & 13.51 & 12.44 \\
\hline
4,32,1 int8  & 28.54 & 18.68 & 9.3 & 4.95 \\
\end{tabular}%
}
\caption{single vs. multi-core performance (ms)}
\label{tab:multi_core}
\end{table}

\section{DFX Methodology}

The DFX methodology is based on the original work by Xilinx \cite{PR} for implementing dynamically reconfigurable systems based on \textit{Pblocks} and black boxes. We extend it to support software-defined systems with encrypted IP cores such as floating-point operators. A \textit{Pblock} defines a floor-planned region associated with a reconfigurable partition, and then each reconfigurable partition has several reconfigurable modules or variants. In this system, the reconfigurable partition has two variants for \textit{int8} and \textit{floating-point}. The black boxes are used to specify the reconfigurable partitions in the static part of the design. In this work, we do not use the DFX Wizard available in Vivado because it  works by selecting the whole IP core as a reconfigurable partition in the Block Diagram. In our case, we work at a finer granularity so we specify reconfigurable regions in RTL generated by the high-level synthesis tool. The idea is to select only the RTL that corresponds the functionally unit that needs to be reconfigured to switch between  \textit{int8}  and \textit{floating-point}.  

\begin{figure}[htb]
  \includegraphics[width=1.0\columnwidth,center]{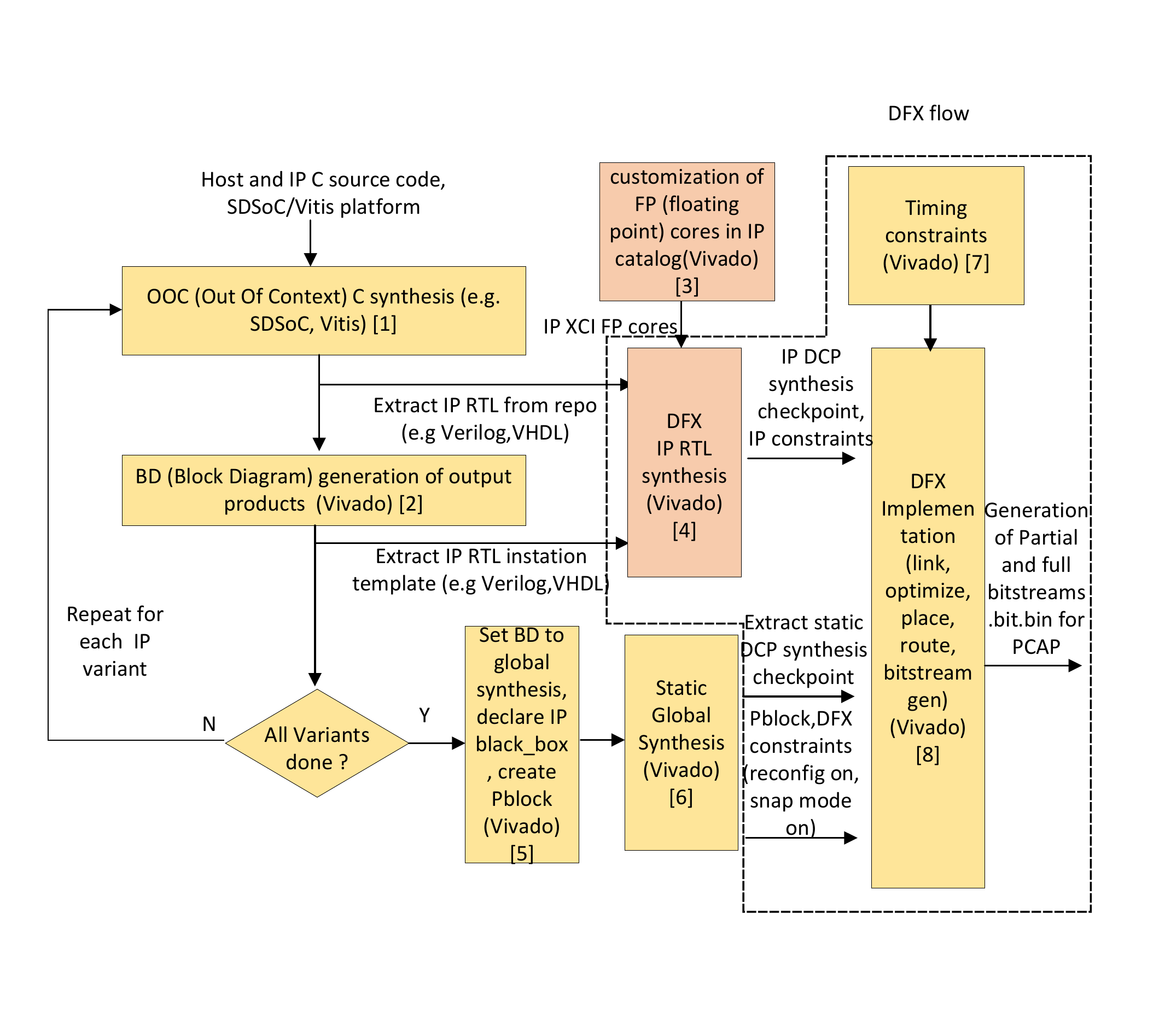}
  \caption{High-level synthesis DFX methodology}
  \label{fig:dfx}
\end{figure}

\begin{figure*}[!htbp]
  \includegraphics[width=17cm,height=7cm]{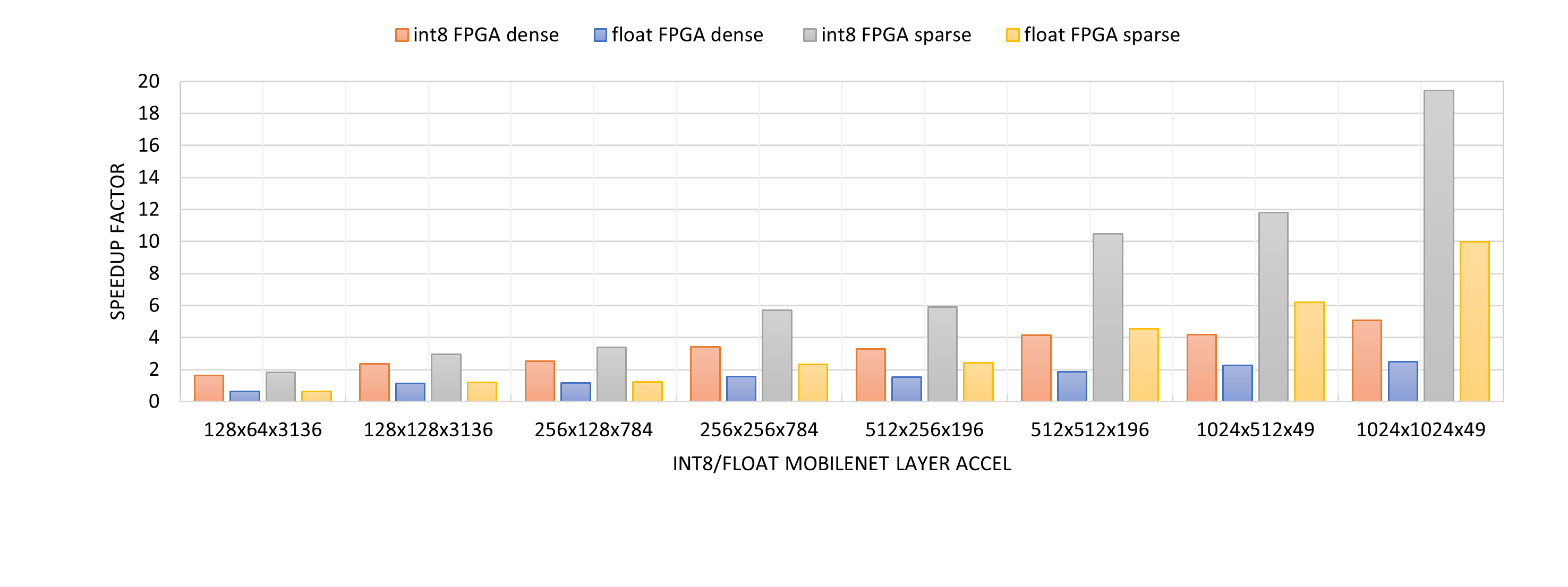}
  \caption{1,32 performance analysis in Tensorflow Lite}
  \label{fig:performance_tflite}
\end{figure*}

A description of the alternative methodology proposed and followed is shown in Fig. \ref{fig:dfx} which shows step numbers described as follows:
\begin{itemize}

\item Step 1: The HLS input code goes through initial C synthesis in Xilinx OOC (Out-of-Context) mode, where each IP core is synthesised independently. The RTL code for the IP is obtained in the corresponding repository. 
\item Step 2: The RTL instantiation template for the IP core is obtained in Vivado after generating output products. The instantiation template and the IP core RTL from the repository are the inputs that define the variations of the IP for the DFX flow (i.e. \textit{int8} or \textit{float}). 
\item Step 3: If this RTL IP core instantiates floating-point cores, they need to be generated using the Vivado IP integrator to avoid black box errors during the subsequent implementation stage. This is because the floating point library is encrypted and it cannot be used directly as an additional RTL file. 
\item Step 4: The standard DFX flow reads the RTL and performs RTL synthesis of each variant of the IP core and generates a DCP checkpoint for each of these variants that are then used by the the rest of the components of the DFX implementation flow. 
\end{itemize}

The steps [5] and [6] in \ref{fig:dfx} correspond to the creation of Pblocks in the floorplanner, declaration of black\_boxes for the dynamic parts of the design and the static synthesis. Finally, timing constraints are defined in [7], and the standard DFX implementation flow [8] places, routes, and generates bitstream files for each of the variants as partial bitstreams in addition to the complete bitstreams resulting from linking the static and dynamic parts of the design. 

The bitstreams with extension .bit.bin files can be used by the processor configuration port (PCAP) available in Zynq devices so the processing system can load complete or partial files. In addition to this hardware flow, the software flow generates a hardware library to invoke hardware execution from the processing system. The API is identical for each of the hardware variants, so a single hardware interface library can be linked with the rest of the software application. To integrate DFX in Tensorflow Lite, we make use of the selective layer quantization feature, possible with Tensorflow 2.7. During inference, the host detects if the layer has scaling parameters available, indicating  the presence of a quantized layer, and consequently checks and configures the programmable logic with the correct IP if necessary. It is essential to maintain the area being reconfigured in an isolated state to avoid unpredictable activity affecting the static part while partial reconfiguration occurs. In this work, we retain the core area and interface under reset by mapping the corresponding clock reset physical registers available in the PS to a userspace area using the Linux memory map function MMAP. This is a simpler alternative than using Partial Reconfiguration (PR) Decoupler IPs to provide logical isolation capabilities for PR designs due to the high number of interface signals in HLS designs.

\section{FADES validation in TFLite} \label{full_perf}
  
In this section we analize the performance, power and energy characteristics after the integration of FADES into the TFLite framework as validation parameters. The core acts as a direct replacement of calls to the RUY library. The TFlite version used is 2.7 and it has been compiled from sources in the Zynq Ultrascale device. In addition to DFX (Dynamic Function Exchange), we consider Virtual Function Exchange (VFX) and Full Reconfiguration (FR) for comparison purposes. In VFX, the hardware implements both \textit{float} and \textit{int8} modes side by side, and multiplexers are used to select which path is active. This virtual mode means that hardware requirements increase but switching between modes can be done in one clock cycle. The hardware complexity of the VFX and DFX modes are compared in table \ref{tab:vfx}.

\begin{table}[!htb]
\resizebox{\columnwidth}{!}{%
\begin{tabular}{llllll}
      Configuration & LUTs(K)  & FFs(K)   & BRAM\_18Ks & DSP48Es  \\
\hline
float DFX  &  90.9 &  120.5 &  186 &  181 \\
\hline
int8  DFX & 50.8 & 56.9 & 189 & 161 \\
\hline
float-int8 VFX & 96.8 & 121.6 & 187.5 & 325 \\
\end{tabular}%
}
\caption{ DFX/VFX hardware complexity comparison}
\label{tab:vfx}
\end{table}

Table \ref{tab:vfx} shows that the additional LUTs/FFs needed in VFX are modest, but the DSP count almost doubles. The additional hardware requirements mean that deploying VFX will limit the number of parallel cores possible and also increase power/energy compared with DFX. The FR mode corresponds to a full bitstream reconfiguration to replace the \textit{int8} hardware with the floating-point hardware. The bitstream size in the considered device is ~25MBytes, and we have measured the time needed to perform this full reconfiguration at ~200ms. On the other hand, the bitstream size for the partial core is ~9MB, and the partial reconfiguration time is around ~30ms. Consequently, there will be close to a factor of 7x higher energy and time overhead using full reconfiguration. In addition to this time advantage of DFX over FR, DFX cores not being reconfigured can continue to operate, so the system does not need to completely hold operation while the full reconfiguration takes place.  

\subsection{FADES performance analysis}


\begin{table*}[htb]
\normalsize
\begin{tabular}{llllllll}
    & ~\cite{efficient} & ~\cite{efficient} & ~\cite{deeper} & ~\cite{deeper} & ~\cite{systolic} & ours x1 & ours x4 \\
\hline
Device  & XCZU9EG & XCZU9EG & XC7Z045 & Nvidia TK1 & XCZU9EG & XCZU9EG & XCZU9EG\\
\hline
Type  & Sparse & Sparse  & Dense & Dense & Dense  & Sparse/Dense & Sparse/Dense\\
\hline
Frequency  & 200 & 200  & 150  & 852 & 200 & 200 & 200\\ 
\hline
Precision  & 16bit & 8bit  & 16bit & float & 8bit & 8bit/float  & 8bit/float\\
\hline
DSPs  & 1350 & 2520  & 780 &  & 1031 & 161/181 & 644/724 \\
\hline
LUTs  & 390k & 405k  & 182k &  & 142k & 50.8K/90.9K & 203k/363k \\
\hline
BRAMs  & 1460  & 1460  & 486 &  & 528 & 189/186  & 700/688 \\
\hline
DSP efficiency\\(Gops/DSP)  & 0.36 & 0.39  & 0.23 &  & 0.18 & 0.47/0.11 & 0.47/0.11 \\
\hline
Performance\\(Gops)   & 495 & 990  & 187  & 76  & 195 & 76/20 & 303/75 \\
\end{tabular}%
\caption{Comparison with other embedded hardware}
\label{tab:embedded_hardware2}
\end{table*}

Figure ~\ref{fig:performance_tflite} shows the performance of a single core 32-PE (1,32) configuration compared with the performance provided by RUY on layers extracted from the Mobilenet neural network. It is clear that larger and wider layers benefit more from the acceleration for both \textit{int8} and float modes, but the hardware can speed up all Mobilenet layers. Configurations with multiple FPGA cores can reduce this execution time significantly, as seen in section \ref{initial} for large layers. The \textit{int8} configuration obtains higher speedups than the floating-point hardware, which can be explained by the good performance of the float RUY software that is not 4x slower than RUY \textit{int8}. RUY uses assembly code to exploit the NEON SIMD accelerator for both integer and float operations. On the other hand, the FPGA float hardware is typically 4x slower than \textit{int8} hardware since each float value contains four \textit{int8} values. The sparse configuration uses a high sparsity level of 90\% and shows significantly better acceleration than the dense configuration. Sparsification is done using the current capabilities of the TensorFlow Model Optimization Toolkit during training and only one layer is sparsified at a time to mimimize the effects on accuracy. We select the layers for pruning in Tensorflow using model cloning with a selective pruning function. We use a polynomial decay strategy and progressively increase pruning from 50\% to the different maximum levels. The paper focuses on showing the hardware capabilities at accelerating sparse layers and not the effects on accuracy of sparsification that is  considered out of scope. The acceleration obtained by the \textit{int8} hardware is higher than the floating-point hardware. It must be noted that thanks to the silicon reuse capability enabled by DFX, it is not necessary to have float and \textit{int8} modes configured simultaneously. This means that configurations with up to 4 cores are possible, resulting in better performance than with the non-reconfigurable alternative (VFX), in which the number of cores should be reduced to accommodate the hardware for supporting both float and \textit{int8} modes.

In Table~\ref{tab:embedded_hardware2} we estimate the performance and hardware details of FADES compared with other embedded hardware available for sparse and dense processing reviewed in section 2. We present results with one core (x1) and four core (x4) configurations and report two values of complexity and performance depending if the active mode is float or int8. 
The table shows that the levels of DSP efficiency are comparable to the best values from the literature while the raw GOPS is proportional to the number of DSP blocks used by the configuration. The main difference and advantage of FADES compared with this hardware is the combination of dense-sparse modes and the switching between int8 and floating-point precision via dynamic reconfiguration. An additional difference is the integration in TFLite with the support of the scaling modes and asymmetric activations part of the TFLite specification. 

\subsection{FADES power and energy Analysis}

The ZCU102 board includes shunt resistors for each of the power rails in the FPGA. These shunts are connected to Texas Instruments INA226 devices that can be used to measure voltage and current. These devices are connected to the I2C bus that is accessible through the \textit{/sys/class/hwmon} interface in the Xilinx Petalinux kernels. We create an application that reads the power values corresponding to the PS and PL and produces a trace of power consumption value, obtaining one power sample per 6 ms due to the limitations of the power measuring setup. 

\begin{figure}[htb]
  \includegraphics[width=1.0\columnwidth,center]{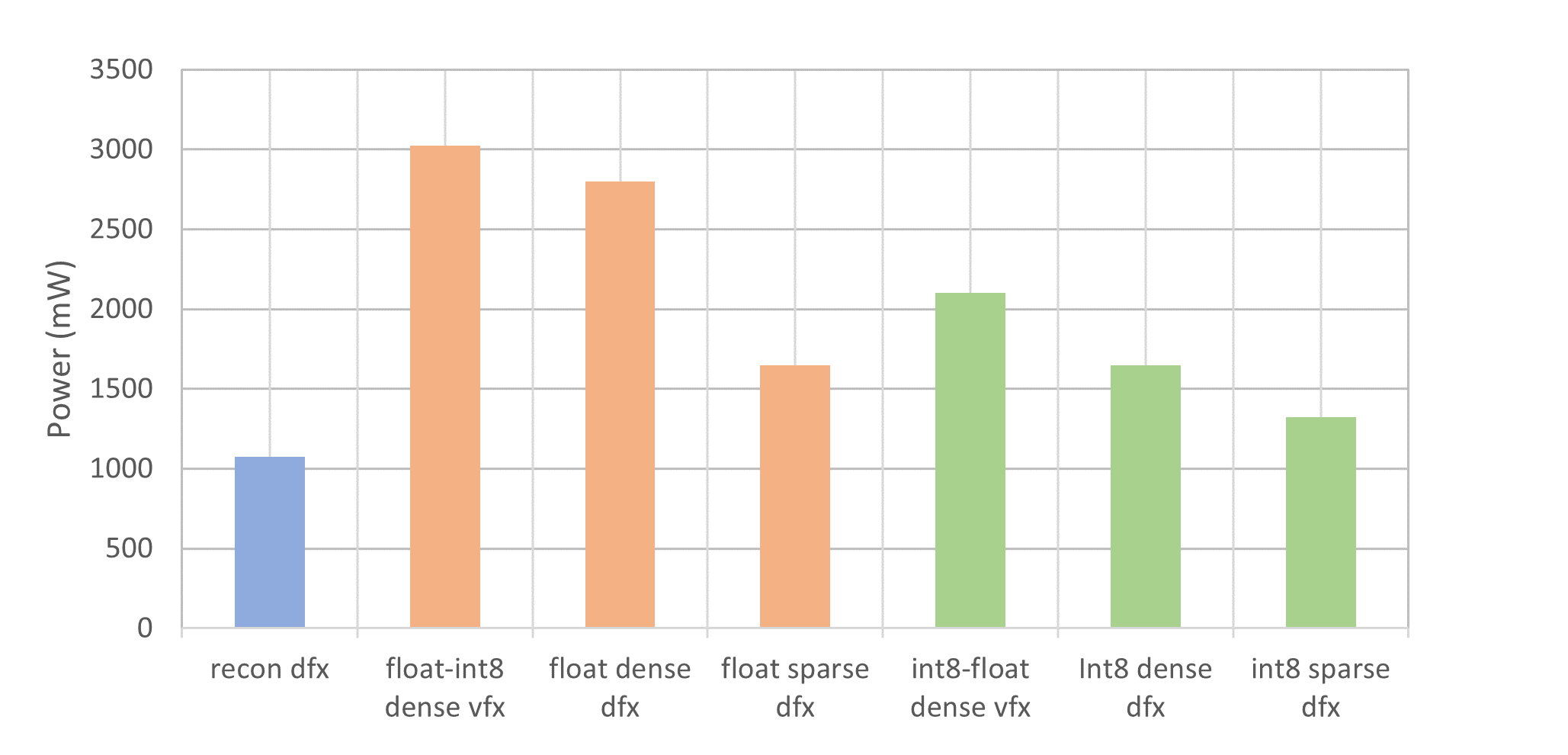}
  \caption{(1,32) power analysis for different configurations}
  \label{fig:power_dfx}
\end{figure}

\begin{figure}[htb]
  \includegraphics[width=1.0\columnwidth,center]{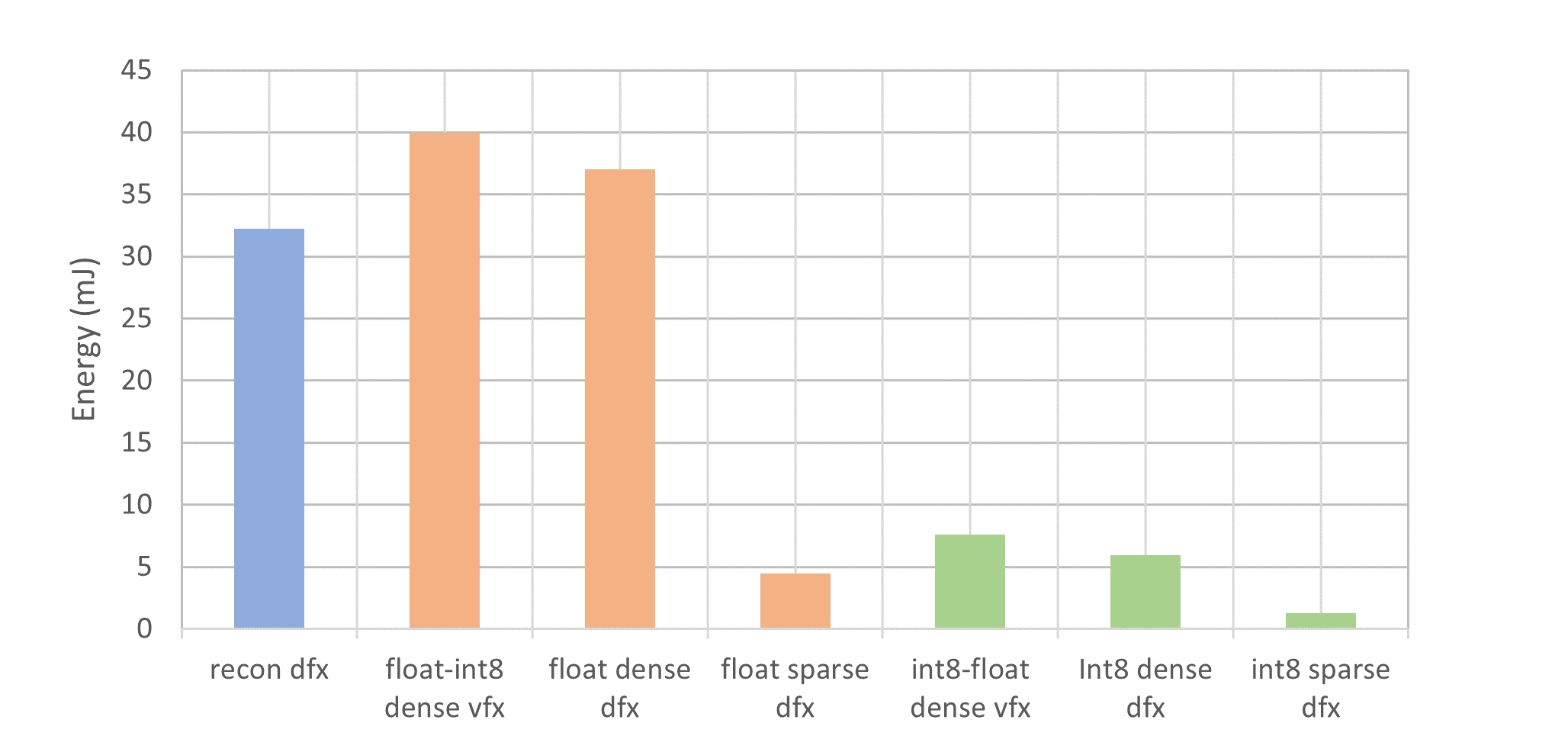}
  \caption{(1,32) energy analysis for different configurations}
  \label{fig:energy_dfx}
\end{figure}

Power usage in the PL is evaluated in Figure ~\ref{fig:power_dfx} that shows that during reconfiguration, power is low since in these runs, no computation is taking place,  and the PS is loading just the configuration memory. PS power is constant mainly during all these runs at around 2 Watts. The most power-intensive modes are the float-int8, and int8-float, which correspond to VFX hardware with both modes configured computing in float or int8 modes, respectively. In DFX, the power of the float mode hardware is higher than the int8 hardware, and this is mainly due to the additional logic used by the pipeline to hide the higher latency of floating-point operations, as seen in Table~\ref{tab:complex}. Note that the additional logic used by the floating-point accelerator enables an equivalent latency of one clock cycle for both int8 and float operands, as seen in section 4.2. There is also a clear decrease of power in the sparse configurations compared to dense. As the amount of sparsity in input $A$ matrix increases, the instantaneous power requirements drop. This can be attributed to the fact that the sparse compute intensity in the PL has a lower number of ops per byte fetched from memory resulting in a lower DSP utilization. 

To verify this DSP utilization, we have estimated the theoretical maximum throughput of the FADES accelerators, assuming 100\% saturation of the core data movement and DSP logic. In the dense mode runs, we measure over 90\% saturation of the data movement and arithmetic pipeline. This utilization reduces as the sparsity level increases with a bottleneck in the data movement that reduces the utilization of the DSP blocks. This can be explained in that although sparsity reduces the amount of data that needs to be moved for the sparse matrix $A$, the dense matrix $B$ still needs to be moved completely, even with a 100\% sparse input. This lower hardware utilization can be observed in the power drop observable in Figure ~\ref{fig:power_dfx}.

Finally, Figure ~\ref{fig:energy_dfx} compares the energy requirements of the Mobilenet layer (1024x1024x49) for different configurations and the dynamic reconfiguration energy cost. The Figure shows that the most energy-intensive configurations are VFX running dense calculations. The DFX dense configurations maintain execution time compared with VFX (at the same number of PEs) but the logic reduction results in lower energy. The sparse DFX configuration benefits from the reduction in execution time and power consumption resulting in the most energy-efficient run. The int8 configuration also benefits from lower power consumption and lower execution time compared with float. The Figure also shows that the energy requirements of the dynamic reconfiguration process between float and int8 hardware are not insignificant, but larger networks with tens of layers would be acceptable if the reconfiguration rate is not very high. For example, when float hardware is only used during training or for groups of layers over time.

\section{Conclusions and Future Work}

The following conclusions can be drawn from the research performed in this paper:

\begin{itemize}
\item Combining sparse and dense modes in a single architecture enables the execution mode selection at a layer level with minimum overheads. Layers can be prepared to run in sparse mode if, during training, it is considered that this mode does not affect accuracy negatively. The sparse mode is faster than the dense mode, even with low sparsity levels of around 50\% thanks to its dataflow architecture that reads CSR values and indices in parallel.
\item Dynamic function exchange can be deployed to optimize the DSP blocks used for the different TFLite \textit{int8} and floating-point precisions. This hardware optimization enables better performance by mapping more cores to a single device, and it also reduces power, avoiding having both pipelines configured simultaneously.  
\item The sparse mode can outperform the dense mode for both float and \textit{int8} precisions, and the dense hardware is also significantly faster than the software optimized dense RUY library and a systolic implementation in the same technology. 
\item The dynamic reconfiguration time has been measured at 30ms which means that it is too slow to enable reconfiguration on a layer-per-layer basis. A more suitable approach would be to deploy dynamic reconfiguration when switching between inference and training modes or switching the network model. 
\item The accelerator needs to buffer a tile of dense matrix with a width that depends on the number of processing elements and a depth that equals the number of rows. For very large matrices there could be not enough BRAM space for all the rows and in these cases tiling at the software level would be needed.  
\end{itemize}

There are several research aspects that can be addressed in the future work:
\begin{itemize}
\item In this research, we have focused on floating point and \textit{int8} precisions because they are natively supported in Tensorflow Lite. Support for 16-bit floats is being introduced, and other research such as  hls4ml~\cite{hls4m} has explored sub-byte precisions such as binary, ternary and quad, although these are not currently part of the TFLite inference engine. DFX can potentially become more valuable to avoid resource overheads as the number of supported arithmetic precisions in TFLite increases.    
\item We have validated the functionality and benefits of DFX using a single core. Further exploring multi-core configurations in which cores not being reconfigured remain active and perform useful computations will be interesting.
\item In the current version the configuration of the accelerator is done manually by the designer. A scheduler that is part of the inference engine running on the host processor and issues reconfiguration commands to the accelerator is needed. This scheduler would need to use a prediction model to decide the best precision and sparsity level for each layer based on accuracy and performance measurements
\item Finally, exploring the application of the FADES accelerator with different precisions for run-time training and other network types such as transformers and recurrent is part of the future work. 
\end{itemize}

\section*{Acknowledgments}

This research was partially supported by the Royal Society Industry fellowship, INF\textbackslash{}R2\textbackslash{}192044 MINET, EPSRC HOPWARE EP\textbackslash{}RV040863\textbackslash{}1, Leverhurme trust international fellowship IF-2021-003 and by the Wallenberg AI autonomous autonomous systems and software (WASP) program funded by the Knut and Alice Wallenberg Foundation.



%

\bibliographystyle{ieeetr}
\bibliography{reference}

%
\begin{IEEEbiography}[{\includegraphics[width=1in,height=1.25in,clip,keepaspectratio]{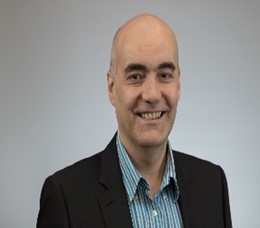}}]{Jose Nunez-Yanez}

Dr Nunez-Yanez is a Professor at Linkoping University in energy efficient and adaptive hardware architectures. Previous to that he was a reader at the University of Bristol.  He holds a PhD in hardware-based parallel data compression from the University of Loughborough. His main area of expertise is in the design of reconfigurable architectures for signal processing with a focus on run-time adaptation, parallelism and energy-efficiency. In 2006-2007 he was a Marie Curie research fellow at STM and in 2011 he was a Royal Society research fellow at ARM Ltd, Cambridge. In 2020-2022 he was an industrial research fellow with the Royal Society at Sensata technologies.
\end{IEEEbiography}

\begin{IEEEbiography}[{\includegraphics[width=1in,height=1.25in,clip,keepaspectratio]{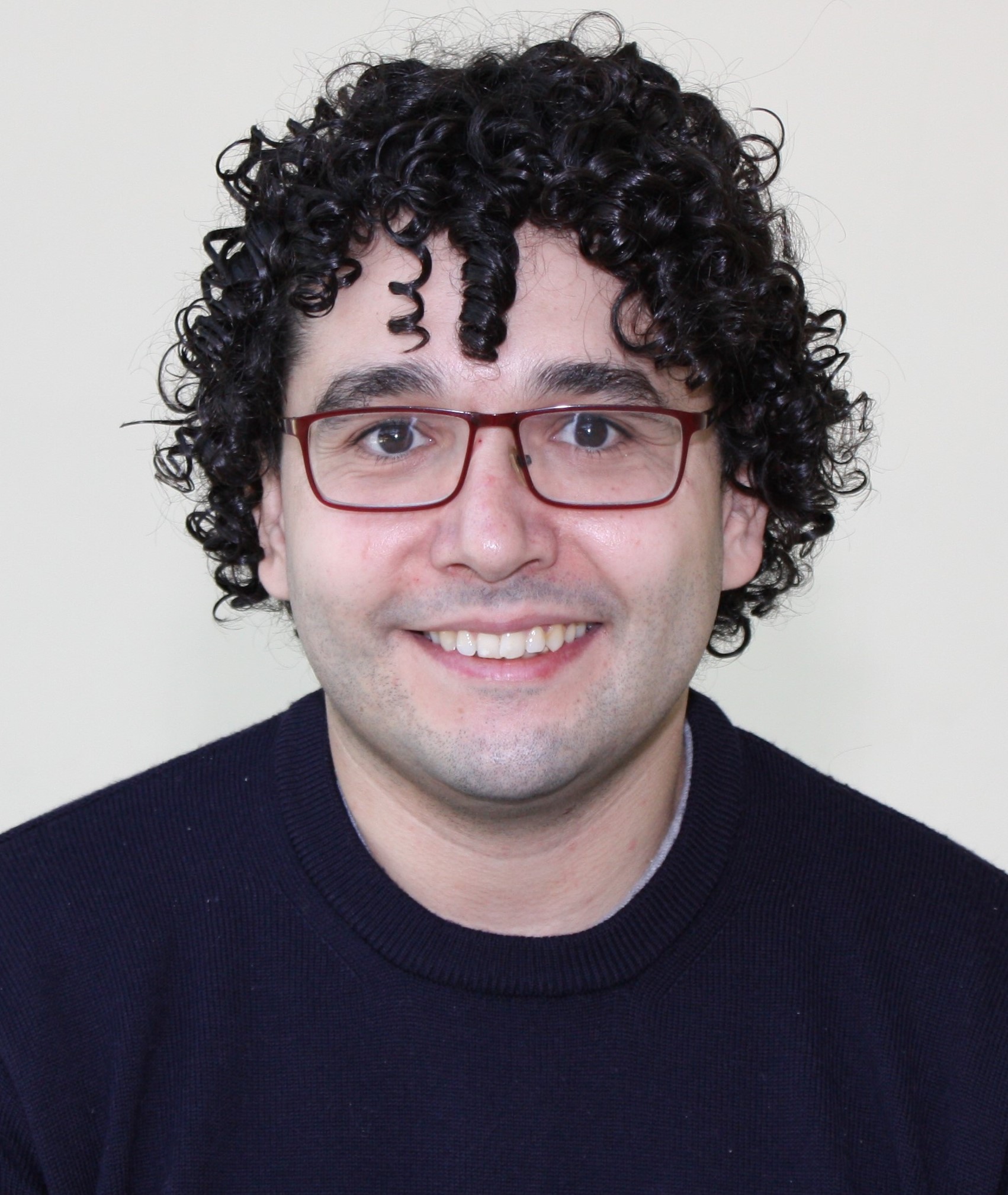}}]{Andr\'es Otero} received his M.Sc. degree in Telecommunication Engineering from the University of Vigo, where he graduated with honors in 2007. He received his Master of Research and Ph.D. degrees in Industrial Electronics from Universidad Polit\'{e}cnica de Madrid (UPM), in 2009 and 2014, respectively. He is currently an Assistant Professor of electronics with the UPM, as well as a researcher in the Centro de Electr\'{o}nica Industrial (CEI). His current research interests are focused on Embedded System Design, Reconfigurable Systems on FPGAs, Evolvable Hardware and Embedded Machine Learning. During the last years, he has been involved in different research projects in these areas, and he is the author of more than 40 papers published in international conferences and journals.
\end{IEEEbiography}
\vskip -2\baselineskip plus -1fil

\begin{IEEEbiography}[{\includegraphics[width=1in,height=1.25in,clip,keepaspectratio]{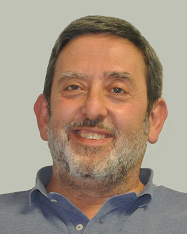}}]{Eduardo de la Torre} is Associate Professor in Electronics at Universidad Polit\'{e}cnica de Madrid (UPM), Spain, doing his research at the Centre of Industrial Electronics. He obtained his PhD in Electrical Engineering from UPM in 2000. His main expertise is in FPGA design, embedded systems design, HW acceleration, signal processing and partial and dynamic reconfiguration of digital systems. He has participated in more than 40 projects, eleven of them being EU funded projects and, overall, in nine funded projects related with reconfigurable systems. 
\end{IEEEbiography}





\end{document}